\documentclass[conference]{IEEEtran}
\usepackage{csquotes}
\usepackage{pifont}
\usepackage{tabularx,booktabs}
\newcolumntype{C}{>{\centering\arraybackslash}X} 
\usepackage[noadjust]{cite}

\ifCLASSINFOpdf
   \usepackage[pdftex]{graphicx}
  \graphicspath{{../figures/}}
  \DeclareGraphicsExtensions{.pdf}
\else
\fi
\usepackage{amsmath}
\usepackage{array}

\ifCLASSOPTIONcompsoc
  \usepackage[caption=false,font=normalsize,labelfont=sf,textfont=sf]{subfig}
\else
  \usepackage[caption=false,font=footnotesize]{subfig}
\fi

\usepackage{stfloats}

\usepackage{url}

\makeatletter
\def\ps@IEEEtitlepagestyle{%
  \def\@oddfoot{\mycopyrightnotice}%
  \def\@evenfoot{}%
}
\def\mycopyrightnotice{%
  \begin{minipage}{\textwidth}
  \centering \scriptsize
  Copyright~\copyright~2024 IEEE.  Personal use of this material is permitted.  Permission from IEEE must be obtained for all other uses, in any current or future media, including\\reprinting/republishing this material for advertising or promotional purposes, creating new collective works, for resale or redistribution to servers or lists, or reuse of any copyrighted component of this work in other works.
  \end{minipage}
}
\makeatother

\begin{document}
%

\title{DTN-COMET: A Comprehensive Operational Metrics Evaluation Toolkit for DTN}

\author{\IEEEauthorblockN{Tobias Nöthlich}
\IEEEauthorblockA{D3TN GmbH\\
Dresden, Germany\\
Email: firstname.lastname@d3tn.com
}
\and
\IEEEauthorblockN{Felix Walter}
\IEEEauthorblockA{D3TN GmbH\\
Dresden, Germany\\
Email: firstname.lastname@d3tn.com}}


%



\maketitle

\IEEEpubidadjcol
\begin{abstract}
Delay- and Disruption-tolerant Networking (DTN) is essential for communication in challenging environments with intermittent connectivity, long delays, and disruptions. Ensuring high performance in these types of networks is crucial because windows for data transmission are sparse and often short. However, research on evaluating DTN implementations is limited. Moreover, existing research relies on manual testing methods that lack reproducibility and scalability. We propose a novel generic framework for reproducible performance evaluation of DTN implementations to address this issue. We validate the framework’s accuracy using a physical testbed and compare the µD3TN and ION DTN implementations. This comparison reveals that µD3TN exhibits higher goodput and shorter bundle retention times. On the other hand, ION exhibited superior memory management and fault tolerance, albeit at the cost of sending and receiving performance. Through this comparison, our framework demonstrates the feasibility of developing a generic toolkit for evaluating DTN.
\end{abstract}


%
\IEEEpeerreviewmaketitle

\section{Introduction}
DTN describes a networking architecture developed for use in challenging environments \cite{Fall03}.
Cerf et al. \cite{rfc4838} describe these environments as characterized by intermittent connectivity, long delays, and link disruptions, as well as potentially high packet loss, which prevents the use of well-established internet protocols such as TCP/IP. 
Because these environments may be highly sensitive, such as networks operated by international space agencies, or battlefield networks in conflict zones, it is crucial that the DTN architecture is stable and performs well.
The importance of performance in DTN is further emphasized by the often limited time available for data transmission between two nodes, which must be utilized effectively.

While the TCP/IP protocol stack provides a robust foundation for the Internet, the DTN architecture, centered around the Bundle Protocol (BP) and its security extension BPsec, is still evolving. 
Although core protocols are standardized, their behavior in diverse network conditions remains relatively unexplored. 
Additionally, critical aspects like network management and key exchange are ongoing research areas. 
These factors complicate the evaluation of DTN implementations. 
Moreover, there is no research on a universal performance evaluation framework suitable for arbitrary DTN implementations.
The absence of such a framework for DTN protocols is a significant gap that hinders progress in the field and practical deployment of DTN solutions.

To our knowledge, this paper is the first to assess the feasibility of realizing a generic performance evaluation framework for DTN implementations.
The main contributions of this paper are:
\begin{itemize}
    \item The first generic and automated performance evaluation framework for DTN implementations.
    \item The first performance comparison of µD3TN and ION.
\end{itemize}

The remainder of the paper is structured as follows: First, we provide some fundamentals on performance evaluation in DTN, including an explanation of the metrics the framework should collect.
A review of the current state of the art follows this.
We then present a detailed concept of the framework.
Section V evaluates and compares µD3TN and ION based on the previously defined metrics.
The paper concludes on the implications for generic DTN toolkit development based on the evaluation results and integration efforts and presents future work.

\section{Background}
For this work, we define performance as the efficiency with which a DTN implementation handles data transmission. This includes the speed of sending and receiving, and the resource usage incurred while processing and generating bundles.

A set of metrics covering important functionalities is required to perform a comprehensive evaluation of DTN performance.
Morgenroth et al. \cite{eval_general} define and explain the influence of key performance metrics on DTN systems and their interdependence. 
Therefore, we will use some metrics introduced in \cite{eval_general} for the performance evaluation framework and introduce new metrics not previously discussed in the literature.

\begingroup
\textbf{Goodput:}\quad 
DTNs are typically operated in environments where data transmission is affected by long delays, sparse contact opportunities, and short contact durations.
Therefore, transmitting as much data as possible during the available connectivity timeframes is crucial.
In this context, goodput has more relevance than throughput, as it describes the data usable by an application rather than including the protocol overhead.
Based on the definition of throughput in \cite{eval_general}, we define the goodput of a DTN implementation as the amount of data usable by an application in Mbit transferred by the DTN implementation in one second.

\textbf{Round-trip time (RTT):}\quad
In DTN, single-trip latency or delivery delay, as described by Morgenroth et al. \cite{eval_general}, is commonly used to measure the time data spends in flight between two nodes due to potential disruptions affecting the round-trip time. 
In the context of the presented testbed, however, there are no delays or disruptions, as the primary objective is to ascertain the maximum achievable performance.
Furthermore, measuring the delivery delay would introduce the issue of clock synchronization between the sending and the receiving node, potentially affecting the accuracy of the results.
Therefore, despite its limited significance in real-world DTN deployments, round-trip time becomes a suitable metric for discovering anomalies during bundle transmission and reception.

\textbf{Memory and CPU usage:}\quad  Because most DTN implementations are intended to run on machines with limited resources such as spacecraft or IoT devices, it is important to consider resource usage during operation.
\endgroup

In addition to the metrics established in related literature, including those we previously presented, we introduce the following metrics that have not yet been discussed in the literature.

\begingroup
\textbf{Bundle retention time (BRT):}\quad We define bundle retention time as the amount of time it takes for a DTN implementation to process a bundle under optimal conditions fully. 
This means that any time overheads incurred during processing are solely due to inefficiencies in the DTN implementation's programming and not due to, for example, waiting for a contact to become active.
This metric allows us to evaluate the average time required for a DTN implementation to process a bundle of a given size, affecting other metrics such as the round-trip time and goodput.

\textbf{(De-)serialization time:}\quad The second new metric we propose is the time a DTN implementation needs for serializing and deserializing a bundle.
Both steps are needed in every DTN implementation and may have a measurable effect on the bundle retention time, allowing for a more detailed breakdown of potential bottlenecks.
\endgroup

\section{Related Work}\label{sec:sota}
The literature in this field primarily focuses on evaluating the performance of a network consisting of interconnected nodes that operate DTN protocol implementations rather than the implementations themselves.
This leads to a focus on routing and forwarding algorithms, and a different set of critical metrics becomes relevant.
Meaningful results in this type of evaluation necessitate many potentially moving nodes. 
In contrast, this paper's primary objective is to assess individual DTN protocol implementations rather than overall network performance. 

This shift redirects the focus from routing and forwarding algorithms to the internal components of DTN protocol implementations, including bundle storage and bundle parsing, as the latter components are more significant in influencing the performance of the DTN implementations themselves. 
Routing and forwarding algorithms, while critical for network performance, are out of this paper's scope, as there already are numerous studies dedicated to investigating their performance.
This leads to a drastic reduction in the number of nodes required for a successful evaluation, as a high number and mobility of the nodes are no longer needed.
Indeed, it negates the primary advantages of simulation- and emulation-based approaches, namely the efficient and cost-effective configuration of numerous nodes. 
Consequently, constructing a physical testbed becomes a viable option. 

Although this kind of performance analysis is essential in ensuring that the frequently short and low-bandwidth connections occurring in DTNs can be used to the fullest of their potential, there is almost no research in creating frameworks performing this sort of evaluation.
There is a gap in the broader field regarding comprehensive performance assessments of DTN protocol implementations.

\begin{table*}
\caption{Overview of Requirements for a DTN Implementation Performance Evaluation Platform Fulfilled by the State of the Art. }
\label{tab:related_works_eval}
    \centering
    \begin{tabularx}{\textwidth}{lCCCCCCCCCCC}
        \cline{1-11}
        && \multicolumn{5}{c}{Collected Metrics (out-of-the-box)}&&&& \\
        \cline{3-7}
        Approach & Auto-mated  & Goodput & RTT & Memory / CPU & BRT & (De)Ser-ialization & Live In-tegration  & Extensi-bility & High \nolinebreak Accuracy & Open-Source\\
\cline{1-11}
        ONE \cite{ONE} &
                \ding{51} & & \ding{51} & & & & & \ding{51} & \ding{51}* & \ding{51}
                \\
        DtnSim \cite{dtnsim} &
                & & \ding{51}† & & & & & \ding{51} & \ding{51}* & \ding{51}
                \\
        ESTNet \cite{estnet} &
                &  & \ding{51} & & &  & & \ding{51} & \ding{51}* & \ding{51}
                \\
        UMass DieselNet \cite{dieselnet} &
                & \ding{51}‡ & \ding{51}‡ &  &  &  & & \ding{51} & \ding{51}* &
                \\
        SPICE \cite{spice} &
                 \ding{51} & \ding{51}‡ & \ding{51}‡ &  & & &  & \ding{51} & \ding{51} &
                \\
        QOMB \cite{qomb} &
                \ding{51}& \ding{51}‡ & \ding{51}‡ & & & &  & \ding{51} & \ding{51} &
                \\
        ION testbed \cite{TCP_ION_Throughput} &
                \ding{51} & \ding{51} & &  & & & & \ding{51} & \ding{51} &
                \\
        TUNIE \cite{TUNIE} &
                \textbf{}& \ding{51} & & & & & & \ding{51} & \ding{51}* &
                \\
        Virtualbricks \cite{Virtualbricks} &
                \ding{51} & \ding{51}‡ & \ding{51}‡ & & & &  & \ding{51} &  & \ding{51}
                \\
        Emustack \cite{emustack} &
                 \ding{51} & \ding{51}‡ & \ding{51}‡ & & &  &  & \ding{51} & &
                \\
        Microbenchmarks \cite{eval} &
                & \ding{51} & & \ding{51} & & & &\ding{51} & \ding{51} &
                \\
        Quantitative analysis \cite{eval_2} &
                & \ding{51} & & & &  & & \ding{51} & \ding{51} &
                \\
        Unified API \cite{unified_api} + DTNperf \cite{dtnperf} &
                 & \ding{51} & \ding{51} & & &   & & \ding{51} & \ding{51} & \ding{51}
                \\
        
        \cline{1-11}
        \multicolumn{11}{l}{* only for routing approaches}\\
        \multicolumn{11}{l}{† delivery delay instead of RTT}\\
        \multicolumn{11}{l}{‡ through applications included with the DTN implementations}\\
        
    \end{tabularx}
    
\end{table*}

Table \ref{tab:related_works_eval} contains an overview over existing simulators (\cite{ONE}-\cite{estnet}), testbeds (\cite{dieselnet}-\cite{TCP_ION_Throughput}), emulators (\cite{TUNIE}, \cite{Virtualbricks}, \cite{emustack}) and manual approaches for DTN implementation performance evaluation (\cite{eval} - \cite{dtnperf}).
It shows whether the proposed toolchains are able to conduct automated tests, comprehensively evaluate the implementations with regards to the previously defined metrics, allow for the integration of new implementations without disrupting the service, are extensible by new metrics, measure accurately, and whether they are open-source.  

Especially simulators often do not fully replicate the properties of physical setups \cite{TUNIE,ONE} or simplify aspects of the DTN architecture \cite{dtnsim}, making it significantly harder to find bottlenecks in the implementations.
The presented testbeds would be able collect these metrics accurately, but there are no frameworks in place which could do so.
Furthermore, the testbeds are not available to the general public and can therefore not be used.
All approaches cannot integrate new DTN implementations without requiring recompilation or definition of new evaluation scenarios.
Furthermore, the results of the studies \cite{eval} and \cite{eval_2} cannot be easily reproduced, as there is no way to repeat the evaluations, which is a core objective of this paper.

\section{Testbed}
Considering the advantages and disadvantages of the different evaluation settings presented in section \ref{sec:sota}, especially the high accuracy of the results obtained on physical testbeds and the elimination of cost efficiency as the primary advantage of simulators and emulators due to the smaller evaluation setup, we decided to build the evaluation framework on a physical testbed. 
We propose a setup consisting of three testbed nodes and one node acting as a Testbed Manager.

\subsection{Testbed Manager}
To provision DTN implementations to the testbed, users can utilize a REST API hosted on the Testbed Manager.
We incorporated a queuing mechanism into the API to prevent potential resource contention between different test runs that may be running simultaneously. This mechanism ensures that only a single test run can be executed at any moment, thus guaranteeing the exclusive use of available resources.
Furthermore, we decided on using SSH via Secure Copy\footnote{\url{https://man7.org/linux/man-pages/man1/scp.1.html}} to provision the archives containing the framework configuration to the testbed nodes, as this approach does not require dedicated services running on the testbed nodes.

\subsection{Testbed nodes}
\begin{figure}[!t]
    \centering
    \includegraphics[width=0.49\textwidth]{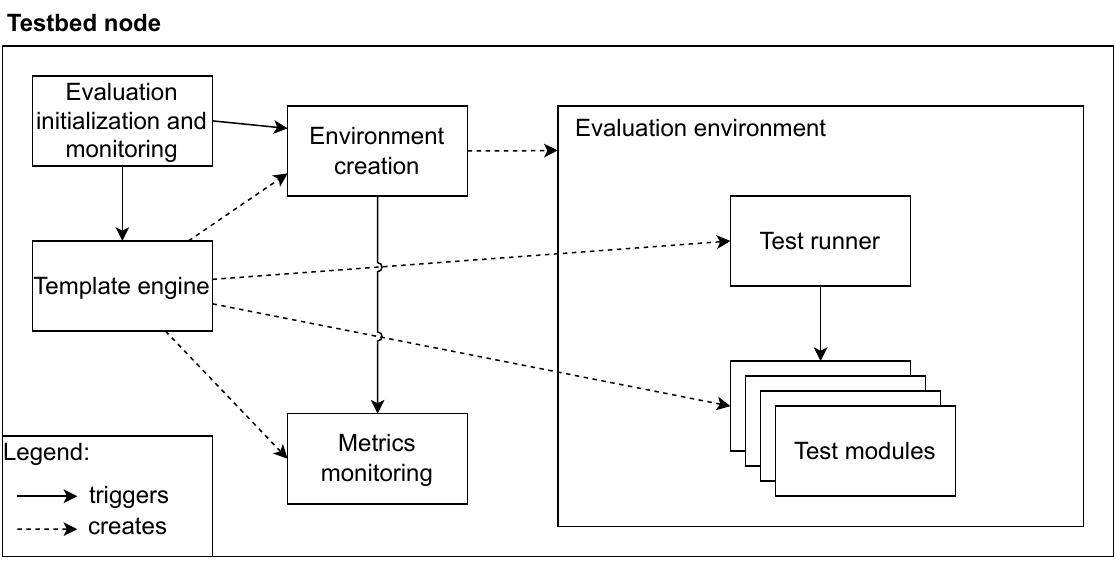}
    \caption{An overview of the modules and their interdependence on the testbed nodes.}
    \label{fig:tb_nodes_modules}
\end{figure}

The framework running on the testbed nodes comprises several scripts and a system service that starts the evaluation workflow upon receiving an evaluation request.
Figure \ref{fig:tb_nodes_modules} presents an overview of its modules and their relation.

The \textbf{evaluation initialization and monitoring module} is responsible for initiating the evaluation process upon receipt of a request from the API node, monitoring the progress of a test run, and performing a clean-up once an evaluation is finished.
This ensures that every test run starts from a clean slate and follows the same steps as previous runs, ensuring repeatability.

The \textbf{template engine} is the first component triggered by the initialization module upon receiving an evaluation request.
This templating process is the key to integrating new DTN implementations.
Most of the scripts involved in setting up and running the test runs contain placeholders, which are substituted with commands and values specified in a framework configuration file supplied by the user.
This ensures that all implementations are tested consistently without manually reconfiguring the testbed before each run.
Developers must create the framework configuration file and implement applications for node pinging, fixed-size bundle transmission, bundle reception with goodput calculation, and variable-payload bundle exchange.
Therefore, since the evaluation is using applications specifically built for the DTN implementation being tested, no modifications to the implementation itself are necessary.
This ensures that DTN implementations with the standard feature set of DTN protocols, even if previously unseen, can be integrated without modifying the framework.

Another module that makes repeatable evaluations possible is the \textbf{environment creation module}.
This module creates Docker containers to keep the DTN implementation under test isolated from the host environment.
Using containerization is not a novel approach, with, e.g., Li et al. \cite{emustack} also leveraging Docker in their emulation platform, as it has many advantages.
Firstly, it addresses potential problems with dependency installation and management.
As users will have to prepare an image containing their implementation and any dependencies, there will be no evaluation failures because of missing or outdated dependencies.
Furthermore, the isolation of the DTN implementations from the host system ensures that side effects that may otherwise impact future evaluations only happen inside a container and do not affect the host directly.

Using container technology to isolate the DTN implementation under test has another benefit.
It becomes significantly easier to monitor the CPU and memory usage of said implementation, as it is the only program running inside the evaluation environment (with the exception of the actual test modules).
This allows us to use a metrics monitoring module to record the memory and CPU usage of the container, which equals the memory and CPU usage of the implementation.

The actual evaluations are performed by a \textbf{test runner} using \textbf{test modules}.
The template engine creates the test runner and modules based on the aforementioned configuration, after which they are copied inside the evaluation environment, and automatically executed.
Having the test broken up into separate modules, which the test runner orchestrates, facilitates the addition of new tests for novel metrics.

\section{Evaluation}
After introducing our generic performance evaluation framework, we use it to evaluate the performance of both the µD3TN (v0.13.0) and ION (v4.1.2) DTN implementations with regard to the previously defined metrics. 

We used the TCP convergence layer adapter v3 \cite{RFC9174} in the tests to ensure comparable results since it is the only CLA that both µD3TN and ION have in common.
Unless otherwise indicated, we have not modified the source code of the two implementations.
However, the following changes have been made to the testbed nodes themselves:
\begin{itemize}
    \item Selective acknowledgments have been activated.
    \item TCP window scaling has been enabled.
    \item The TCP send and receive buffer sizes have been increased, to a maximum of 12 MByte.
    \item The network interface's receive queue size has been increased to 5000 frames.
\end{itemize}

These modifications increased the mean goodput measured using iperf3 to 638 MBit/s, with recorded values ranging from 581 MBit/s to 708 MBit/s.

For ION, we had to modify the default configuration slightly for a fair comparison.
The new configuration (in each node's \emph{.ionconfig} file) disables the \textsc{SDR\_REVERSIBLE} and \textsc{SDR\_BOUNDED} configuration flags set by default and increases the size of the memory ION allocates for its SDR database to 40 MByte.
This ensures the node does not run out of memory during the goodput tests.
By setting \textsc{SDR\_REVERSIBLE} to false, we disable reversible transactions in ION, as their impact on performance is significant.
Furthermore, disabling \textsc{SDR\_BOUNDED} removes the restriction on SDR heap updates crossing object boundaries.
No further configuration is required for µD3TN except for the general setup of the convergence layers and contacts used during the test, which also applies to ION.

The evaluations were run on a testbed consisting of 3 ZUBoard 1CG MPSoC development boards\footnote{\url{https://www.avnet.com/wps/portal/us/products/avnet-boards/avnet-board-families/zuboard-1cg/}}, featuring a dual-core Arm Cortex-A53 MPCore, 1 GB LPDDR4 RAM, and Gigabit Ethernet.

\subsection{Round-trip Times}
\begin{figure}[!t]
    \centering
    \includegraphics[width=0.49\textwidth]{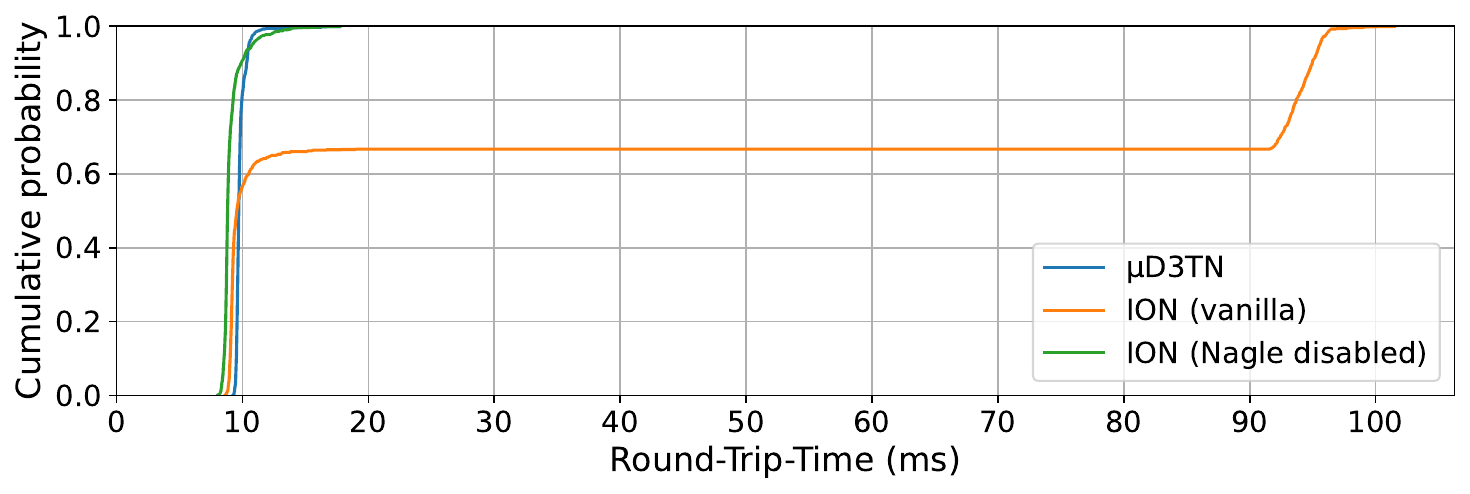}
    \caption{Cumulative probabilities of round-trip times per implementation for 1000 bundles.}
    \label{fig:rtt_timeseries}
\end{figure}

Figure \ref{fig:rtt_timeseries} shows an ECDF plot of the cumulative probabilities of round-trip times for the bundles sent by each implementation.
There is a significant outlier for the unmodified ION implementation, in which only about 66\% of bundles can be expected to have a round-trip time of below 15 milliseconds.
This can be explained by examining the source code of ION.
When creating the sockets for sending bundles, ION does not set the \textsc{TCP\_NODELAY} socket option.
This activates Nagle's algorithm for the socket, which holds back every third bundle to prevent flooding the network with many small packets.
However, this method of improving TCP efficiency is not desired in this case.
Therefore, we modified the ION source to set the \textsc{TCP\_NODELAY} option when the socket is created.
As shown in Figure \ref{fig:rtt_timeseries} this resulted in more consistent round-trip times, similar to the round-trip times achieved by µD3TN.

\subsection{Goodputs}
We performed goodput measurements for both versions of ION, even though the bundle sizes should be large enough to avoid triggering Nagle's algorithm. 

\begin{figure}[!t]
    \centering
    \includegraphics[width=0.49\textwidth]{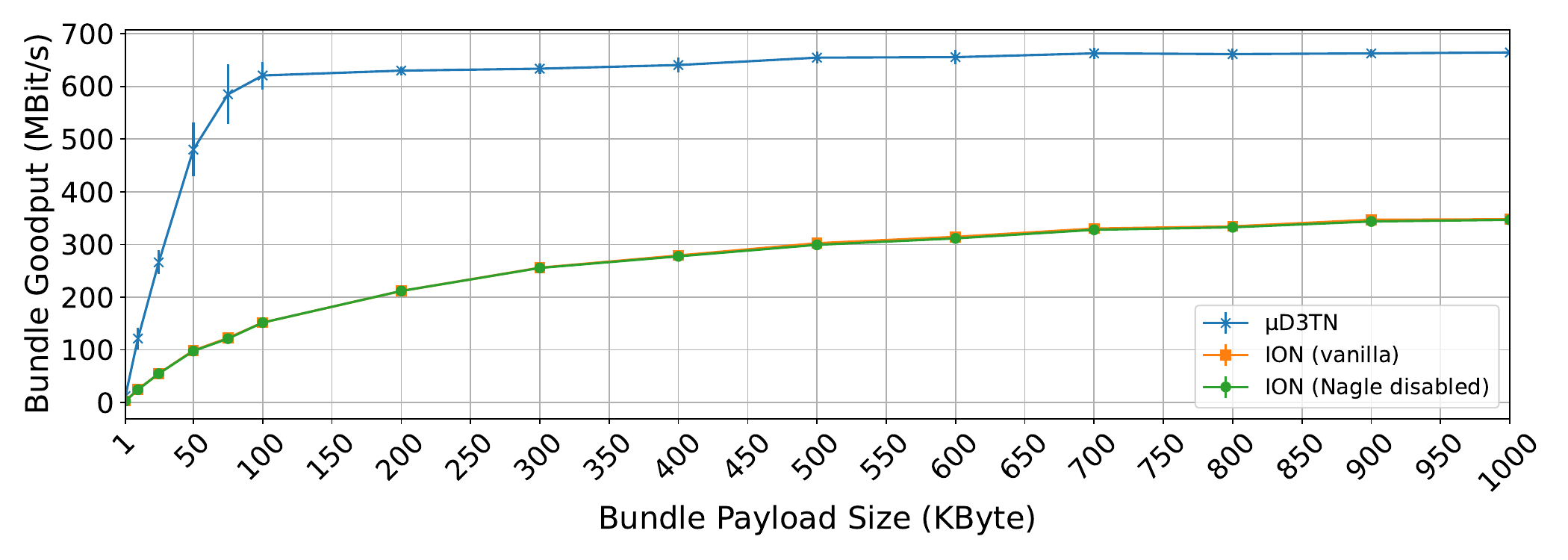}
    \caption{Average goodput for different bundle payload sizes including 99\% confidence intervals. For each payload size 10,000 bundles were sent.}
    \label{fig:throughput}
\end{figure}

Figure \ref{fig:throughput} displays the goodputs achieved by the implementations for a given bundle payload size.
The overhead associated with bundle creation, storage, and forwarding for small bundles greatly limits the achievable goodput. 
This is particularly noticeable when examining the achieved goodputs for bundle payload sizes less than 50 KByte.

Overall, µD3TN achieves significantly higher goodput than ION, reaching the maximum possible goodput on the testbed.
This is true for all measurements with payloads larger than 100 KByte, effectively capping the maximum goodput that µD3TN can achieve.
This efficiency is likely a result of the bundle processing in µD3TN, where all bundles are stored in RAM. 
This eliminates the need for expensive read and write operations and completely obviates the need to interact with any form of external storage.
Interestingly there is a lot of variability in the recorded goodputs for smaller bundle sizes up to 100 KByte, which indicates that there are some variable overheads especially affecting the processing of bundles. 

Conversely, ION achieves a maximum goodput of roughly 350 MBit/s, which is about half of µD3TN's.
This outcome is not surprising, as the processing in ION is significantly more complex than in µD3TN.
Additionally, the bottleneck in these goodput measurements lies on the receiving side in ION, as attested by Huff \cite{TCP_ION_Throughput}, which means that µD3TN processes bundles faster, resulting in higher goodputs compared to ION.
However, we can also see that the confidence intervals in both ION tests are significantly smaller than the ones calculated for µD3TN, which implies that bundle processing overheads are less variable in ION.
A comparison of the different versions of ION shows that Nagle's algorithm had no notable effect on the measured goodput.

\subsection{CPU and Memory Usage}
Both of these metrics are used to indicate whether a given DTN implementation can run efficiently on embedded systems.
However, evaluating the average CPU usage is not a comparison but rather a check to determine whether a DTN implementation frequently exceeds a given threshold. 
Comparing the values of different DTN implementations would not be fair due to the inclusion of the load caused by applications that create bundles for the DTN implementations, such as in the goodput test.
For instance, µD3TN primarily uses Python scripts, whereas ION relies on C programs, which results in lower CPU usage.

\begin{figure}[!t]
    \centering
    \includegraphics[width=0.49\textwidth]{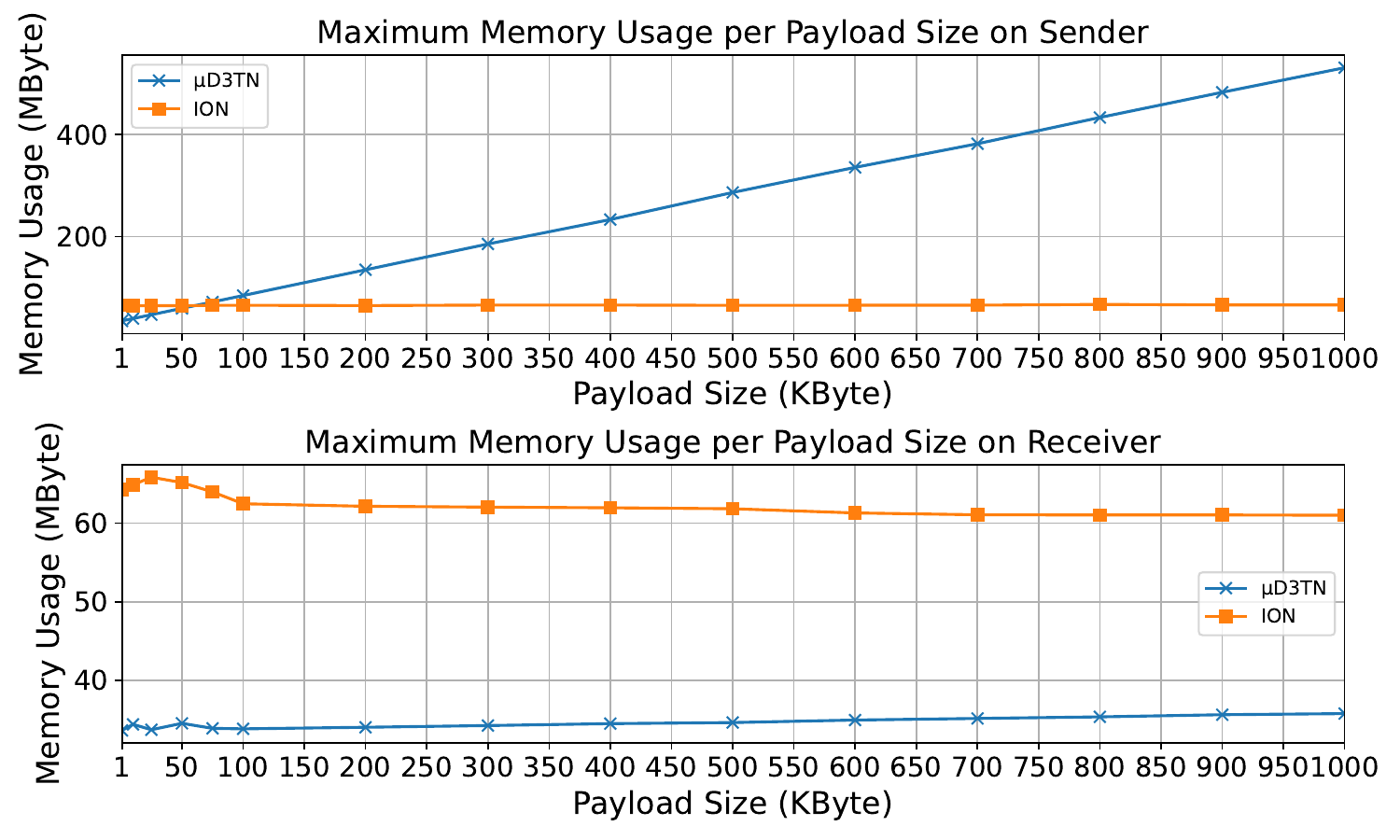}
    \caption{Measured memory usage for different payload sizes. Note the fluctuations in recorded memory usage for ION and µD3TN due to the inclusion of page cache in the measurements.}
    \label{fig:memory}
\end{figure}

Figure \ref{fig:memory} shows the maximum memory usage of µD3TN and ION for different bundle payload sizes during the goodput tests on both the sending and receiving nodes.

On the sending node, we observed that µD3TN's memory usage increased linearly, while ION's memory usage remained nearly constant.
This can be explained by the differences in architecture between the two implementations.
As µD3TN v0.13.0 stores all bundles in RAM, an increase in bundle payload size or number of bundles sent results in a proportional increase in memory usage.
In contrast, ION allocates a fixed amount of dynamic memory during initialization and does not reallocate, resize or release it until the ION node is terminated.
Most of the data from a bundle is stored in the filesystem, with no impact on memory usage.

The memory measurement accuracy depends on the allocation method chosen by the implementation. 
For example, ION uses the \emph{shmget()} system call to create a System V shared memory segment, which resides completely in the kernel page cache.
However, this predominant use of the page cache may impact the accuracy of the memory usage measurement. This is because ION is not the only application to use the cache, and therefore usage may fluctuate in an unpredictable manner. 

Table \ref{tab:cpu} presents the key statistics on CPU usage for the sender and receiver nodes.
From this data it can be inferred that both implementations are suitable for use on resource-constrained systems, with µD3TN and ION averaging 5\%-18\% and 6\%-10\% CPU usage on a single core, respectively.
The table also shows that the third quartile of CPU utilization for both implementations is well below 10\% (µD3TN: 5\%, ION: 2\%-3\%). 
This indicates that there is sufficient headroom to handle potential performance spikes.
The CPU usage statistics further reveal that generating and sending bundles is the most CPU-intensive task for µD3TN, while the opposite is true for ION.
This is consistent with Huff's \cite{TCP_ION_Throughput} findings, which also indicate that reception speed is the limiting factor in ION if both nodes have identical specifications.

\begin{table}
    \centering
    \caption{Summary statistics of the CPU usage.}
    \label{tab:cpu}
\begin{tabular}{lcc|cc}
\toprule
& \multicolumn{2}{c}{\textbf{µD3TN}} & \multicolumn{2}{c}{\textbf{ION}} \\
\midrule
Metric & Sender & Receiver & Sender & Receiver \\
\midrule
Mean &
    18\% & 5\% & 6\% & 10\%\\
Standard deviation &
    35\% & 16\% & 12\% & 27\% \\
25\% &
    0\% & 0\% & 0\% & 0\% \\
50\% &
    0\% & 0\% & 0\% & 0\% \\
75\% &
    5\% & 0\% & 3\% & 2\% \\
\bottomrule
\end{tabular}
    
\end{table}

\subsection{Bundle Retention Times}

\begin{figure}[!t]
\centering
    \subfloat[ION]{\includegraphics[width=0.25\textwidth]{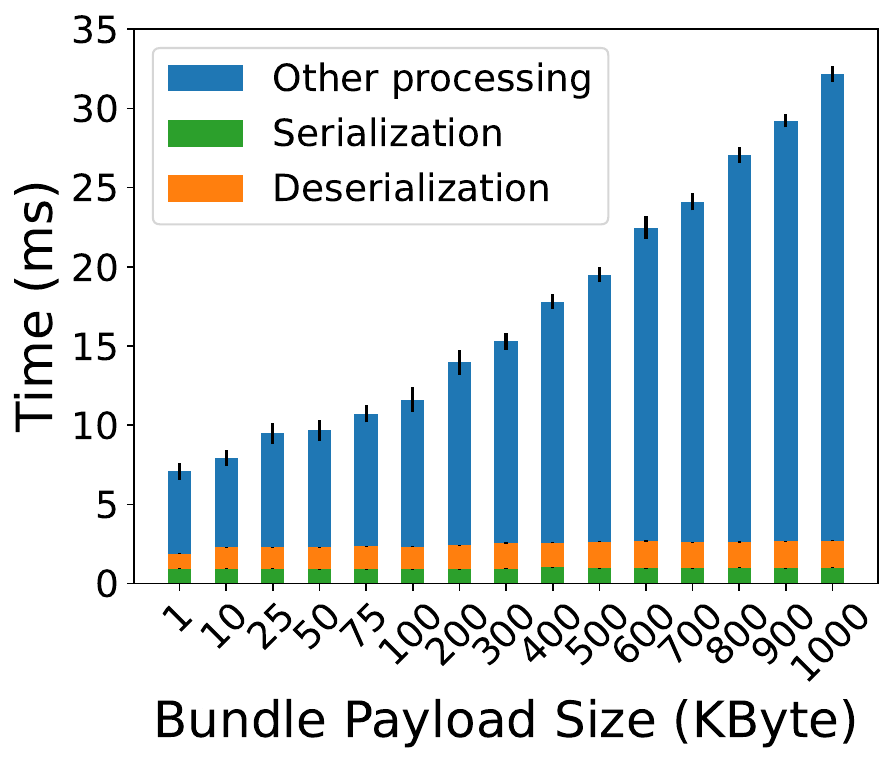}\label{fig:brt_ion}}%
    \subfloat[µD3TN]{\includegraphics[width=0.25\textwidth]{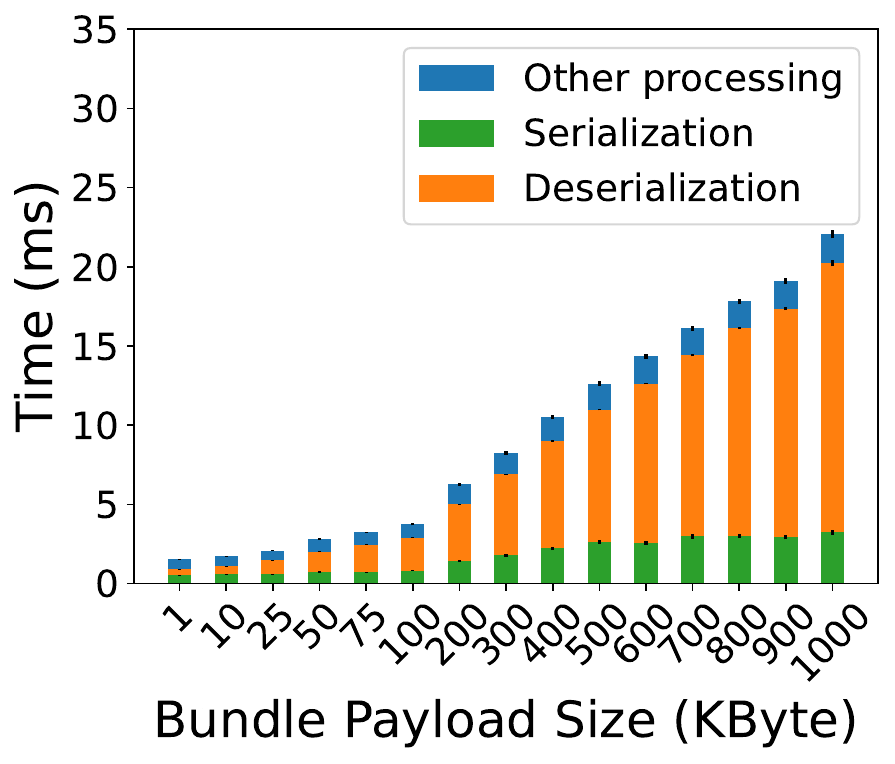}\label{fig:brt_ud3tn}}
    \caption{Average bundle retention, serialization, and deserialization times measured for ION and µD3TN. For each payload size 1,000 bundles were sent.}%
    \label{fig:brts}%
\end{figure}

Figure \ref{fig:brts} compares the average bundle retention times of µD3TN and ION based on the bundle payload size.
The data indicates that the retention time of a bundle in µD3TN is generally shorter than in ION.

The figure also contains the times for serialization and deserialization of a bundle overlaid on the total bundle retention time.
This effectively segments the data into the time spent on (de-)serialization and other processing, such as routing, queuing, and writing or reading the bundle data.
Comparing the two figures, an interesting observation can be made: The time spent on tasks that are not related to (de-)serialization in ION increases as the payload size increases, whereas in µD3TN, the time spent on processing that does not involve serializing or deserializing a bundle remains mostly constant.

The cause of this issue can be found in how the DTN implementations handle bundle reception and parsing.
Parsing in µD3TN occurs as data is received.
However, this process also includes the \emph{recv()} syscall for copying the bundle data from the kernel buffer into the user space as part of the deserialization process. The sharp increase in deserialization time is a result of these copying steps.

The approach utilized in ION differs slightly as the deserialization step happens after receiving the entire bundle rather than using a streaming approach like µD3TN.
After receiving the bundle, parsing is done in a single function call using a zero-copy object that contains the bundle's bytes.
The time-consuming \emph{recv()} syscall is therefore not included in the deserialization time but instead considered to be a part of the \emph{other processing}.

\section{Conclusion}
In this paper, we present a novel framework for evaluating the performance of DTN implementations.
In addition, a physical test environment was created using realistic hardware specifically for executing this framework.
While the current framework can evaluate DTN implementations, its functionality could be further extended by incorporating link modeling capabilities (including disruptions) and reworking the Testbed Manager to allow for interoperability testing. 
This would facilitate the collection of novel metrics and enable the development of more comprehensive evaluation scenarios, particularly those focused on assessing the interoperability performance between different DTN implementations deployed on separate nodes.

We found that a critical challenge in developing generic DTN toolkits is the lack of standardization across implementations. 
This leads to inconsistent interactions, as differences in exposed interfaces and design choices complicate integration. 
This issue becomes particularly clear in the context of application registration, which is handled in various ways across different implementations.



\section*{Acknowledgment}
The authors would like to thank the team at D3TN for the stimulating discussions and their feedback on the proposed framework. 
The framework was developed as part of the D3TN-ES project co-funded by the European Union.

\bibliography{bibliography}{}
\bibliographystyle{IEEEtran}

\end{document}